# Meditations on Quantified Constraint Satisfaction

Hubie Chen

Departament de Tecnologies de la Informació i les Comunicacions
Universitat Pompeu Fabra
Barcelona, Spain
`hubie.chen@upf.edu`

**Abstract.** The quantified constraint satisfaction problem (QCSP) is the problem of deciding, given a structure and a first-order prenex sentence whose quantifier-free part is the conjunction of atoms, whether or not the sentence holds on the structure. One obtains a family of problems by defining, for each structure **B**, the problem QCSP(**B**) to be the QCSP where the structure is fixed to be **B**. In this article, we offer a viewpoint on the research program of understanding the complexity of the problems QCSP(**B**) on finite structures. In particular, we propose and discuss a group of conjectures; throughout, we attempt to place the conjectures in relation to existing results and to emphasize open issues and potential research directions.

*This article is dedicated to Dexter Kozen on the occasion of his 60th birthday. Congratulations and thanks, Dexter, for all that you have given us.*

## 1 Introduction

The constraint satisfaction problem (CSP) is a general computational problem that involves deciding, given a set of constraints on a set of variables, whether or not there is an assignment to the variables satisfying all of the constraints. Problems from many areas of computer science can be formulated within this general framework, including graph coloring problems, boolean satisfiability problems, and problems from temporal and spatial reasoning. The CSP can be formalized logically as the problem of deciding, given a prenex $\{\exists, \wedge\}$-sentence $\Phi$ and a structure **B**, whether or not $\Phi$ holds on **B**. By a $\{\exists, \wedge\}$-sentence, we mean a first-order sentence built from atoms, the connective $\wedge$, and existential quantification. Deciding such a sentence in prenex form amounts to deciding if there is an assignment to the variables (each of which is existentially quantified) that satisfies all of the atoms, which can be thought of as constraints on the variables. The quantified constraint satisfaction problem (QCSP) is the natural generalization of the CSP in which universal quantification is also permitted: it is the problem of deciding, given a prenex $\{\forall, \exists, \wedge\}$-sentence $\Phi$ and a structure **B**, whether or not $\Phi$ holds on **B**. The addition of the universal quantifier permits the expression of a wider range of problems, including problems from quantified propositional logic and game theory. The greater expressiveness of the QCSP comes at the cost of higher computational complexity: if one considers finite structures, the CSP is in general NP-complete, while the QCSP is in general PSPACE-complete.

For each of these two problems, one can obtain restricted cases of interest by considering the problem relative to a fixed structure. Formally, for each structure **B**, define CSP(**B**) (respectively, QCSP(**B**)) to be the problem of deciding, given a $\{\exists, \wedge\}$-sentence $\Phi$ (respectively, $\{\forall, \exists, \wedge\}$-sentence $\Phi$), whether or not $\Phi$ holds on **B**. As examples, consider 2-SAT and Horn-SAT, two well-known polynomial-time tractable cases of the boolean satisfiability problem. One can capture these two problems as problems of the form CSP(**B**), by defining appropriate structures **B**; correspondingly, the quantified generalizations of these problems, Quantified 2-SAT and Quantified Horn-SAT, can each be formulated as a problem of the form QCSP(**B**).

A research direction with roots in seminal articles by Schaefer [31] and Feder and Vardi [19] is to understand the complexity behavior of the problem family CSP(**B**) on finite structures **B**. Schaefer [31] classified the complexity of CSP(**B**) on two-element structures: he described the two-element structures **B** such that CSP(**B**) is in P, showing that for all other two-element structures **B**, the problem CSP(**B**) is NP-complete. Feder and Vardi [19] studied the problems CSP(**B**) on general finite structures, and famously conjectured that this problem family admits a dichotomy in that each problem therein is either in P or is NP-complete. In the 90s, Jeavons and co-authors pioneered an algebraic approach to studying the problems CSP(**B**), which involves associating an algebra to each finite structure **B**, and then using properties of the algebra of **B** to gain insight into and derive results on the complexity of CSP(**B**); see for example [24,23,22,8]. Although this article's focus is on finite structures, it can be remarked that this algebraic approach has been developed by Bodirsky and co-authors for $\omega$-categorical structures [5,4,3].

The algebraic approach to studying the family CSP(**B**) has been the subject of focused research in the 00s which continues to the present (see for example [9,26,27,1,21] and the references therein). As the set of tools for understanding this family has developed, researchers have also studied many variants of the CSP with the aim of giving full complexity classifications on all finite structures **B**. One example variant is the QCSP; as another, one can name the *counting CSP*, wherein one wants to count the number of satisfying assignments, in place of deciding if one exists [10]. (See [25,18,7] for further examples.) Complexity classification results typically give broad sufficient conditions for tractability, intractability, or (more generally) completeness for a complexity class; they can often be used as the basis for analyzing the complexity of further problems.

In this article, we offer a viewpoint on the research direction of understanding the complexity of the problem family QCSP(**B**) over finite structures. In particular, we propose and discuss a group of conjectures concerning the complexity of the problems QCSP(**B**). Throughout our presentation, we attempt to place the conjectures in relation to existing results, and to emphasize open issues and po-tential research directions. One of the leading protagonists of these conjectures is a property of algebras called the *polynomially generated powers (PGP) property*: essentially, an algebra $\mathbb{A}$ has this property when its powers $\mathbb{A}, \mathbb{A}^2, \mathbb{A}^3, \ldots$ have generating sets of polynomial size. This property is a close relative of the *few subpowers property*, another property of algebras that concerns a growth rate in the powers of an algebra and which has given insight into the CSP [2,21]; see [17] for a discussion. We point out that in the case of the CSP, a precise conjecture predicting the boundary of the P/NP-complete dichotomy was posed in the early 00s, in the conference version of [8]; inside P, a picture of the situation has emerged with demonstrations of general hardness criteria for sub-P complexity classes [27].

## 2 Preliminaries

We study relational first-order logic. A *signature* is a set of relation symbols, each of which has an associated arity. A *relational structure* $\mathbf{B}$ over a signature $\sigma$ consists of a set $B$ called the *universe* and a relation $R^{\mathbf{B}} \subseteq B^k$ for each relation symbol $R \in \sigma$; here, $k$ denotes the arity of $R$. We will frequently use the symbol $\mathbf{B}$ to denote a relational structure with universe $B$. In this article, we focus on finite relational structures, by which is meant structures having finite universes. Let $\mathbf{B}$ be a structure on signature $\sigma$. We use $\mathbf{B}^*$ to denote the expansion of $\mathbf{B}$ containing all constant relations; precisely, $\mathbf{B}^*$ is a structure over signature $\sigma^* \supseteq \sigma$ with universe $B$ such that (1) for each $b \in B$, there exists a relation symbol $R_b \in \sigma^*$ of arity 1 where $R_b^{\mathbf{B}^*} = \{b\}$, and (2) each relation symbol in $\sigma^* \setminus \sigma$ is of the form $R_b$ where $R_b^{\mathbf{B}^*} = \{b\}$ and such that for no other relation symbol $S \in \sigma^*$ does it hold that $S^{\mathbf{B}^*} = \{b\}$. Note that we have $(\mathbf{B}^*)^* = \mathbf{B}^*$.

Let $\sigma$ be a signature. We define a *quantified constraint formula* over $\sigma$ to be a first-order formula of the form $Q_1 v_1 \ldots Q_n v_n \phi$ where each $Q_i \in \{\forall, \exists\}$ is a quantifier, each $v_i$ is a variable, and $\phi$ is the conjunction of $\sigma$-predicate applications. By a $\sigma$-predicate application, we mean a formula $R(w_1, \ldots, w_k)$ where $R \in \sigma$, the $w_i$ are variables, and $k$ is the arity of $R$. We define a *constraint formula* to be a quantified conjunctive formula that does not make use of universal quantification, that is, where all quantifiers are existential. We will use the term *(quantified) constraint sentence* to refer to a (quantified) constraint formula having no free variables.

The CSP (QCSP) is the problem of deciding, given a (quantified) constraint sentence $\Phi$ and a structure $\mathbf{B}$, whether or not $\mathbf{B} \models \Phi$. Each structure gives rise to a restricted version of each of these problems, defined as follows. For a structure $\mathbf{B}$, the problem $\mathsf{CSP}(\mathbf{B})$ is the problem of deciding, given a constraint sentence $\Phi$, whether or not $\mathbf{B} \models \Phi$; similarly, the problem $\mathsf{QCSP}(\mathbf{B})$ is the problem of deciding, given a quantified constraint sentence $\Phi$, whether or not $\mathbf{B} \models \Phi$. We will be interested in studying the QCSP where, in addition to fixing the structure, the number of quantifier alternations is bounded. We define a representative sequence of problems as follows. For each $k \geq 1$, we define $\Pi_{2k}\text{-}\mathsf{QCSP}(\mathbf{B})$ to be the restriction of the problem $\mathsf{QCSP}(\mathbf{B})$ to quantifier prefixes that are $\Pi_{2k}$. Note that the only notion of complexity-theoretic reduction that we will make use of in this paper is many-one polynomial-time reduction.

Let $B$ be a nonempty set, let $f : B^n \to B$ be an $n$-ary operation on $B$, and let $R \subseteq B^k$ be a $k$-ary relation on $B$. We say that $f$ is a *polymorphism* of $R$ (and that $f$ *preserves* $R$) if for every length $n$ sequence of tuples $t_1, \ldots, t_n \in R$, denoting the tuple $t_i$ by $(t_{i,1}, \ldots, t_{i,k})$, it holds that the tuple

$$f(t_1, \ldots, t_n) = (f(t_{1,1}, \ldots, t_{n,1}), \ldots, f(t_{1,k}, \ldots, t_{n,k}))$$

is in $R$. We extend this terminology to relational structures, and say that an operation $f$ is a polymorphism of a relational structure $\mathbf{B}$ if $f$ is a polymorphism of every relation of $\mathbf{B}$.

For our purposes, an *algebra* $\mathbb{A}$ is a pair $(A, F)$ consisting of a non-empty set $A$ called the *universe* and a set $F$ of finitary operations on $A$. Let $\mathbb{A} = (A, F)$ be an algebra. A *subalgebra* of $\mathbb{A}$ is an algebra of the form $(B, F|_B)$, where $B \subseteq A$ and $B$ is preserved by all operations in $F$. By $F|_B$, we mean the set $\{f|_B \mid f \in F\}$, where $f|_B$ denotes the restriction of the operation $f$ to the set $B$. The subalgebra (of $\mathbb{A}$) *generated* by a subset $X \subseteq A$ is defined to be the intersection of all $\mathbb{A}$-subalgebras containing $X$. We say that a subset $X \subseteq A$ generates an algebra $\mathbb{A} = (A, F)$ if $\mathbb{A}$ itself is the subalgebra of $\mathbb{A}$ generated by $X$. We will frequently use the symbol $\mathbb{A}$ to denote an algebra $(A, F)$.

Our focus in this paper is on finite structures of the form $\mathbf{B}^*$, that is, finite structures having a relation for each constant. We thus define $\mathbb{A}_\mathbf{B}$, the algebra for structure $\mathbf{B}$, to be the algebra $(B, \mathsf{IPol}(\mathbf{B}))$ where $\mathsf{IPol}(\mathbf{B})$ denotes the set of all idempotent polymorphisms of $\mathbf{B}$. Recall that an operation $f : B^n \to B$ is idempotent if for all $b \in B$, it holds that $f(b, \ldots, b) = b$. Note that for any structure $\mathbf{B}$, it holds that $\mathsf{IPol}(\mathbf{B}) = \mathsf{IPol}(\mathbf{B}^*)$ and hence $\mathbb{A}_\mathbf{B} = \mathbb{A}_{\mathbf{B}^*}$. For a finite structure $\mathbf{B}^*$, the set of idempotent polymorphisms $\mathsf{IPol}(\mathbf{B}^*)$ can be connected to the complexity of $\mathsf{QCSP}(\mathbf{B}^*)$; see [6].

In the case of the CSP, it is known that for each finite structure $\mathbf{A}$, there exists a finite structure $\mathbf{B}$ such that the problems $\mathsf{CSP}(\mathbf{A})$, $\mathsf{CSP}(\mathbf{B}^*)$ are polynomial-time interreducible [8]. In the context of CSP complexity classification, this result justifies focusing on structures having constants, that is, structures of the form $\mathbf{B}^*$. In contrast, for the QCSP, no such passage to structures having constants is known. From this author's perspective, any results contributing to an understanding of whether or not such a passage exists would be highly appreciated. Perhaps the discussion in the last section of this article will provide some clues!

## 3  The Polynomially Generated Powers Property

For an algebra $\mathbb{A}$, let $d(\mathbb{A})$ denote the smallest size of a generating set for $\mathbb{A}$. The sequence $(d(\mathbb{A}), d(\mathbb{A}^2), d(\mathbb{A}^3), \ldots)$ is called the *d-sequence* of an algebra $\mathbb{A}$. Study of the $d$-sequence goes back to work of Wiegold and co-authors [32,33,34,35,29,36], which focused on groups and semigroups. Recent years have seen a revival of interest in the $d$-sequences of algebras; see for example [30,20]. Previous work by the present author [17] connected the complexity of $\mathsf{QCSP}(\mathbf{B})$ with polynomially bounded growth of the $d$-sequence of $\mathbb{A}_\mathbf{B}$; this mode of growth is formalized as follows.

**Definition 1.** *An algebra $\mathbb{A}$ has the* polynomially generated powers (PGP) *property if there exists a polynomial $p(n)$ (on the natural numbers) such that for all $n \geq 1$, there exists a subset $X_n \subseteq A^n$ of size $|X_n| \leq p(n)$ that generates the algebra $\mathbb{A}^n$.*

In particular, one can derive implications for the QCSP when the polymorphism algebra has the PGP property in an effective sense; effectiveness of this property is formalized as follows.

**Definition 2.** *An algebra $\mathbb{A}$ has the* effective PGP *property if there exists an algorithm that, given a natural number $n \geq 1$, outputs in polynomial time (in $n$) a subset $X_n \subseteq A^n$ that generates the algebra $\mathbb{A}^n$.*

*Remark 1.* We will discuss the effective PGP property only on finite algebras. We require that the algorithm in the definition outputs each set $X_n$ in the form of an explicit listing of $n$-tuples. Since in polynomial time it is only possible to output polynomially many tuples, the effective PGP property implies the PGP property.

*Example 1.* Consider an algebra $\mathbb{A}$ with universe $B$ having a binary operation $f : B^2 \to B$ with respect to which there is an identity element, that is, such that there exists $e \in B$ such that for all $b \in B$, it holds that $f(b, e) = f(e, b) = b$. We will argue that such an algebra $\mathbb{A}$ has the PGP. Monoids and groups are examples of algebras having this property.

For each $n \geq 1$, for each $i \in \{1, \ldots, n\}$, and for each $b \in B$, define $t[n, i, b]$ to be the tuple in $B^n$ whose $i$th entry is equal to $b$, and which has all other entries equal to $e$. For each $n \geq 1$, define $T_n$ to be the set

$$\{t[n, i, b] \mid i \in \{1, \ldots, n\}, b \in B\}.$$

We claim that, with respect to the algebra $\mathbb{A}$, the set $T_n$ generates $B^n$. Let $(b_1, \ldots, b_n)$ be an arbitrary element of $B^n$. By $f$-multiplying together the tuples $t[n, 1, b_1], t[n, 2, b_2], \ldots, t[n, n, b_n]$ in any order, one arrives at $(b_1, \ldots, b_n)$, establishing the claim. We have $|T_n| \leq n|B|$, that is, the size of the sets $T_n$ exhibits linear growth, and hence the algebra $\mathbb{A}$ has the PGP with respect to the polynomial $p(n) = n|B|$. Moreover, one can readily devise a polynomial-time algorithm that, given $n$, outputs $T_n$, and thus the algebra $\mathbb{A}$ has the effective PGP.

Recall that a *semilattice operation* is a binary operation that is associative, commutative, and idempotent. On the Boolean universe $\{0, 1\}$, there are two semilattice operations, the Boolean AND ($\wedge$), which has identity element 1, and the Boolean OR ($\vee$), which has identity element 0. Consider the 3-element algebra $\mathbb{A} = (\{a, b, c\}, \{f\})$ where $f : \{a, b, c\}^2 \to \{a, b, c\}$ is defined by the following rule: $f(x, y) = x$ if $x = y$, and $f(x, y) = c$ otherwise. The operation $f$ is a semilattice operation. It can be verified that, for any generating set $T_n$ of $\mathbb{A}^n$, it holds that $T_n \supseteq \{a, b\}^n$; this follows from the fact that if $f$ is applied to two

$n$-tuples $s, s'$ and the result is a tuple $t \in \{a, b\}^n$, then $s = s' = t$. Consequently, any such generating set $T_n$ must have $|T_n| \geq 2^n$, and this 3-element algebra $\mathbb{A}$ does not have the PGP. More generally, it can be shown that if a semilattice operation $f : B^2 \to B$ does not have an identity element, then for an appropriate choice of distinct elements $a, b \in B$ one again has that any generating set of $B^n$ must contain $\{a, b\}^n$, and thus that the algebra $(B, \{f\})$ does not have the PGP. □

The following theorem directly connects the QCSP to the effective PGP property. Essentially, it states that when the algebra $\mathbb{A}_\mathbf{B}$ has this property, the bounded-alternation QCSP on $\mathbf{B}$ can be reduced to $\mathsf{CSP}(\mathbf{B}^*)$ and is hence in NP.

**Theorem 1.** *[17] Let $\mathbf{B}$ be a finite relational structure on a finite signature. If the algebra $\mathbb{A}_\mathbf{B}$ has the effective PGP property, then for all $k \geq 1$, the problem $\Pi_{2k}\text{-}\mathsf{QCSP}(\mathbf{B})$ reduces to $\mathsf{CSP}(\mathbf{B}^*)$ and is in NP.*

We give an explanation of how this theorem is proved.

*Proof idea of Theorem 1.* It suffices to prove the result under the assumption that $\mathbf{B} = \mathbf{B}^*$, which is what we do. The proof is an induction argument showing that, for all $k \geq 1$, there exists a polynomial-time algorithm that converts a $\Pi_{2k}$ quantified constraint formula $\Phi$ to a constraint formula $\Phi'$ that is equivalent in the natural sense: a $B$-assignment $f$ to the free variables satisfies $\Phi$ over $\mathbf{B}$ if and only if it satisfies $\Phi'$ over $\mathbf{B}$.

For $k = 1$, this is done as follows. Let $Y$ denote the universally quantified variables in $\Phi$. A generating set $\{t_1, \ldots, t_m\}$ for $\mathbb{A}_\mathbf{B}^{|Y|}$ is computed in polynomial time using the algorithm giving the effective PGP property; each element $t_i$ in the set can (and will!) be naturally viewed as a mapping from $Y$ to $B$. For each $i$ from 1 to $m$, a formula $\Phi_i$ is created from $\Phi$ by changing all universal quantifiers to existential quantifiers, and then conjoining $\bigwedge_{y \in Y} R_{t_i(y)}(y)$ to the quantifier-free part. In the formula $\Phi_i$, the universally quantified variables (of $\Phi$) are forced to be set according to the generator $t_i$. The desired formula $\Phi'$ is (the prenexing of) the formula $\bigwedge_{i=1}^m \Phi_i$. The point is that to determine the truth of $\Phi$ (relative to $\mathbf{B}$ and an assignment to the free variables), in lieu of considering *all* possible assignments to the universally quantified variables $Y$, one can consider just the assignments $t_i$ from the generating set. (We note that verifying this makes use of the idempotence of the operations of $\mathbb{A}_\mathbf{B}$.) In essence, having a polynomially-sized generating set allows us to short-cut the consideration of the exponentially many assignments to the universally quantified variables.

The induction is argued as follows. Suppose that one has the result for $k$; we want to show the result for $k + 1$. Given a $\Pi_{2(k+1)}$ quantified constraint formula $\Phi$, by removing the outermost two quantifier blocks, we obtain a $\Pi_{2k}$ formula $\Phi_2$. We use the algorithm for $\Pi_{2k}$ formulas, which exists by induction, to obtain an equivalent constraint formula. Taking this constraint formula and putting back the two quantifier blocks that were removed, we obtain a $\Pi_2$ formula, which can then be handled using the algorithm presented for the case $k = 1$. The overall algorithm in this case, modulo some simple syntactic manipulations, is the composition of two polynomial-time algorithms, and is hence itself polynomial-time. □

The argument just given certainly allows one (under the assumption of the effective PGP property) to translate any instance $\Phi$ of $\mathsf{QCSP}(\mathbf{B})$ to a truth-equivalent instance of $\mathsf{CSP}(\mathbf{B}^*)$: simply determine the lowest $k$ such that $\Phi$ is $\Pi_{2k}$, and then apply the algorithm given for $\Pi_{2k}$ formulas. However, it is worth noting that the action of this translation on *all* instances of $\mathsf{QCSP}(\mathbf{B})$ is *not* in polynomial time. We give a brief explanation as to why. Consider an algebra $\mathbb{A}$ that has no generating set of size 1, that is, where every generating set has size greater than or equal to 2. Let us examine how the translation acts on a $\Pi_{2n}$ formula of the form $\forall y_1 \exists x_1 \ldots \forall y_n \exists x_n \phi$. The translation works from the inside out; first, the subformula $\forall y_n \exists x_n \phi$ is converted to a constraint formula $\phi_1$. The resulting constraint formula arises as a conjunction (of copies of $\phi$) over a generating set for $\mathbb{A}$, and so by assumption has size at least 2. Then, the translation converts the formula $\forall y_{n-1} \exists x_{n-1} \phi_1$ to a constraint formula $\phi_2$; again, the result is a conjunction of copies of $\phi_1$ over a generating set for $\mathbb{A}$, and so has size at least 4. Continuing to argue in this way, one obtains that the size of each intermediate formula $\phi_i$ has size at least $2^i$, and the final constraint formula $\phi_n$ has size at least $2^n$.

Our first conjecture is that each finite algebra has either the PGP property or has a $d$-sequence with exponential growth; the latter property is formalized as follows.

**Definition 3.** *An algebra $\mathbb{A}$ has the* exponentially generated powers (EGP) property *if there exists a real number $b > 1$ such that the mapping (on the positive natural numbers) that takes $n$ to $d(\mathbb{A}^n)$ is $\Omega(b^n)$.*

Note that a dichotomy between the PGP property and the EGP property cannot be taken for granted, as there are growth rates (such as $n^{\log n}$) that are neither polynomial nor exponential, according to the definitions here.

*Conjecture 1.* Each finite algebra $\mathbb{A}$ either has the PGP property or has the EGP property.

Support for this conjecture can be found, for example, in the article of Wiegold [36] on semigroups, and in previous work relating the PGP property to the QCSP, where a class of 3-element algebras was studied [17].

A parenthesis. We will, later in this paper, conjecture that absence of the PGP property leads to coNP-hardness of the QCSP (see Conjecture 3 and Remark 4). Intuitively speaking, existing coNP-hardness results in this vein (which sometimes apply to structures having a tractable CSP) use a block of universally quantified variables to induce the consideration of an exponentially large search space, in such a way that the quantified constraint sentence is true if and only if there is no object of desirable type in the search space. Conjecture 1 is algebraic: we are not aware of any direct implications that it would have for QCSP complexity. But, could belief in the computational Conjecture 3 support belief in the algebraic Conjecture 1?

## 4  Bounded Alternation

In this section, we present and discuss three conjectures in the setting of bounded alternation. The first we view as quite innocuous: it states that, on finite algebras, the PGP property is always effective. The truth of this conjecture would allow us to directly connect the PGP property to the bounded-alternation QCSP.

*Conjecture 2.* For every finite algebra $\mathbb{A}$, if $\mathbb{A}$ has the PGP property, then $\mathbb{A}$ has the effective PGP property.

*Remark 2.* From a proof of Conjecture 2, one would be able to simplify the statement of Theorem 1: one could remove the effectiveness assumption on the PGP property.

*Remark 3.* Previous work on the QCSP [17] identified a property on algebras called *switchability* which implies the PGP; in fact, a look at the generating sets given by switchability allows one to readily verify that switchability implies the effective PGP. Within a particular family of 3-element algebras, it was shown that the PGP implies switchability, providing evidence for Conjecture 2. We believe that it could be of interest to investigate whether or not all algebras having the PGP property are switchable.

So, Conjecture 2 predicts that the presence of the PGP property places the bounded-alternation QCSP in NP. Our next conjecture predicts that the PGP property is, in fact, the only explanation for this QCSP being in NP, on structures having the form $\mathbf{B}^*$.

*Conjecture 3.* Let $\mathbf{B}$ be a finite relational structure on a finite signature. If the algebra $\mathbb{A}_\mathbf{B}$ does not have the PGP property, then there exists $k \geq 1$ such that $\Pi_{2k}\text{-QCSP}(\mathbf{B}^*)$ is coNP-hard.

*Remark 4.* Examples of coNP-hardness results for the QCSP which provide evidence for Conjecture 3 can be found in the articles [11,12,13].

The two conjectures given in this section (thus far) predict when the bounded-alternation QCSP on structures $\mathbf{B}^*$ will be in NP. Indeed, these conjectures predict that if the sequence $\Pi_{2k}\text{-QCSP}(\mathbf{B}^*)$ is in NP, then the algebra $\mathbb{A}_\mathbf{B} = \mathbb{A}_{\mathbf{B}^*}$ has the PGP and each problem in this sequence reduces to (and is hence equivalent to) $\text{CSP}(\mathbf{B}^*)$, as in Theorem 1. Outside of NP, one can observe two different modes of complexity behavior (for the QCSP under bounded alternation). One is coNP-completeness. An example of this behavior is given by certain structures preserved by semilattice operations without an identity element, for which one has, for each $k \geq 1$, coNP-completeness of $\Pi_{2k}\text{-QCSP}(\mathbf{B}^*)$: coNP-hardness comes from [11], and containment in coNP follows from results in [15]. Another behavior is that the complexity increases unboundedly with the prefix class; this happens already in the two-element case, see [16, Section 7].

We believe that it should also be of interest to study when the bounded-alternation QCSP can be placed in coNP. We have the following conjecture concerning this matter.

*Conjecture 4.* Let $\mathbf{B}$ be a finite relational structure on a finite signature. If the problem $\mathsf{CSP}(\mathbf{B}^*)$ is in P, then for all $k \geq 1$, the problem $\Pi_{2k}$-$\mathsf{QCSP}(\mathbf{B}^*)$ is in coNP.

*Remark 5.* A previous article [15] gives coNP inclusion results for the bounded alternation QCSP in a model that is more general than $\Pi_{2k}$-$\mathsf{QCSP}(\mathbf{B}^*)$ in that the restriction to the structure is applied only to the existentially quantified variables. The technology developed there should be of help in trying to establish Conjecture 4. The coNP inclusion results presented therein that apply to structures of the form $\mathbf{B}^*$, such as the results on structures preserved by semilattice operations, apply directly to the problems $\Pi_{2k}$-$\mathsf{QCSP}(\mathbf{B}^*)$ and give evidence for Conjecture 4.

For a structure $\mathbf{B}^*$, when neither the NP upper bound predicted by Conjecture 2 nor the coNP upper bound predicted by Conjecture 4 applies, we believe that the complexity of the problems $\Pi_{2k}$-$\mathsf{QCSP}(\mathbf{B}^*)$ should increase unboundedly; perhaps in this case, for each $k \geq 1$, one has $\Pi_{2k}^p$-completeness of $\Pi_{2k}$-$\mathsf{QCSP}(\mathbf{B}^*)$, a behavior that can be observed in the case of two-element structures [16, Section 7].

## 5  Unbounded Alternation

In previous articles [14,17], with the motivation of giving positive QCSP complexity results, we identified and studied properties of algebras called *collapsibility* and *switchability*; switchability is a generalization of collapsibility. Each of these properties implies both the effective PGP property and a reduction from the QCSP to the CSP akin to that given by Theorem 1, but in the general setting of unbounded alternation. Intuitively, the reductions exploit the particular form of generating sets that give the PGP property. Based on this work and as an extension of Conjecture 2 (with Theorem 1), we propose that the PGP property implies a QCSP-to-CSP reduction under unbounded alternation.

*Conjecture 5.* Let $\mathbf{B}$ be a finite relational structure on a finite signature. If the algebra $\mathbb{A}_\mathbf{B}$ has the PGP property, then $\mathsf{QCSP}(\mathbf{B})$ reduces to $\mathsf{CSP}(\mathbf{B}^*)$, and $\mathsf{QCSP}(\mathbf{B})$ is in NP.

*Remark 6.* Since $\mathbb{A}_\mathbf{B} = \mathbb{A}_{\mathbf{B}^*}$ and $\mathsf{QCSP}(\mathbf{B})$ trivially reduces to $\mathsf{QCSP}(\mathbf{B}^*)$, in order to prove Conjecture 5, it suffices to verify it on structures of the form $\mathbf{B}^*$.

As examples showing how the PGP property can be linked to the QCSP complexity properties of Conjecture 5, we revisit two classes of structures whose QCSP is already known to be in NP [28]. We show that each structure therein has an algebra that is *collapsible*, from which it follows that the algebra has the PGP property and the structure enjoys a QCSP-to-CSP reduction [17].

*Example 2.* (Bipartite graphs) Let $\sigma = \{E\}$ be the signature containing a single binary relation. We consider a structure $\mathbf{B}$ on this signature to be a *bipartite*

*graph* if $E^{\mathbf{B}}$ is symmetric and there exists a partition $B = B_0 \cup B_1$ of the universe $B$ of $\mathbf{B}$ such that $E^{\mathbf{B}} \subseteq (B_0 \times B_1) \cup (B_1 \times B_0)$. It is known that for each bipartite graph $\mathbf{B}$, the problem $\mathsf{QCSP}(\mathbf{B})$ is in P [28].

Let $w : \{0,1\}^5 \to \{0,1\}$ be the operation that, given a 0/1 tuple of length 5, outputs the unique element in $\{0,1\}$ that occurs 3 or more times in the tuple; the operation $w$ can be thought of as returning the "majority winner", given a tuple of 5 votes. Let $\mathsf{mid}$ be the operation defined on non-empty subsets of $\{1,2,3,4,5\}$ that, given a set of size $n$, ranks the elements in the set as $a_1 < \cdots < a_n$, and outputs the element $a_{\lceil \frac{n}{2} \rceil}$. Intuitively, the operation $\mathsf{mid}$ outputs the "middle" element in a given set according to the natural $<$-ranking; on sets of even cardinality, the lower of the two elements in the middle is preferred. For example, $\mathsf{mid}(\{1,4,5\}) = 4$ and $\mathsf{mid}(\{1,2,4,5\}) = 2$.

We now define an operation $p : B^5 \to B$, relative to a partition $B = B_0 \cup B_1$, as follows. Let $s : B \to \{0,1\}$ be the mapping that, given $b \in B$, indicates which side of the partition $B_0 \cup B_1$ the element $b$ lies in; that is, $s$ is defined to be the unique mapping such that for all $b \in B$, it holds that $b \in B_{s(b)}$. Define $p : B^5 \to B$ to be the operation $p(b_1, \ldots, b_5) = b_{p'(b_1,\ldots,b_5)}$, where

$$p'(b_1, \ldots, b_5) = \mathsf{mid}(\{i \mid s(b_i) = w(s(b_1), \ldots, s(b_5))\}).$$

In words, the operation $p$ computes the set of coordinates among $\{1, \ldots, 5\}$ that support the majority winner of the given elements (under $s$); it then takes the middle coordinate in that set, and projects the given tuple onto that coordinate. The operation $p$ is clearly idempotent.

We show that $p$ is a polymorphism of any bipartite graph $\mathbf{B}$. Suppose that $(b_1, b'_1), \ldots, (b_5, b'_5) \in E^{\mathbf{B}}$. We want to show that $(p(b_1, \ldots, b_5), p(b'_1, \ldots, b'_5)) \in E^{\mathbf{B}}$. We claim that $p'(b_1, \ldots, b_5) = p'(b'_1, \ldots, b'_5)$, which suffices. Since $\mathbf{B}$ is bipartite, for each $i$ (ranging from 1 to 5), we have $s(b_i) \neq s(b'_i)$. It follows that $w(s(b_1), \ldots, s(b_5)) \neq w(s(b'_1), \ldots, s(b'_5))$ and that

$$\{i \mid s(b_i) = w(s(b_1), \ldots, s(b_5))\} = \{i \mid s(b'_i) = w(s(b'_1), \ldots, s(b'_5))\},$$

from which the claim follows.

Collapsibility is a property on algebras which was introduced in the study of the QCSP; collapsibility of an algebra $\mathbb{A}_{\mathbf{B}}$ implies a reduction from $\mathsf{QCSP}(\mathbf{B})$ to $\mathsf{CSP}(\mathbf{B}^*)$ [14]. For an operation $f : A^k \to A$, a coordinate $i \in \{1, \ldots, k\}$, and an element $a \in A$, let $f_i^a : A^{k-1} \to A$ be the operation obtained by $f$ by fixing the $i$th argument of $f$ to be the element $a$. It has been shown [14, Lemma 5.13] that an algebra is collapsible if it has an idempotent operation $f : A^k \to A$ and there exists an element $a \in A$ such that each of the $k$ operations $f_i^a$ is surjective. We use this fact to prove the collapsibility of $\mathbb{A}_{\mathbf{B}}$ when $\mathbf{B}$ is a bipartite graph, thus showing that for any bipartite graph $\mathbf{B}$, it holds that $\mathsf{QCSP}(\mathbf{B})$ reduces to $\mathsf{CSP}(\mathbf{B}^*)$. Fix an element $c \in B_0$. We show that each of the 5 operations $p_1^c, \ldots, p_5^c$ is surjective; by the cited lemma, this gives the desired reduction.

We first consider the case $i \neq 3$; in this case, we claim that $p_i^c$ is idempotent, that is, for all $b \in B$, it holds that $p_i^c(b, b, b, b) = b$. The value $p_i^c(b, b, b, b)$ is equal to the value of $p$ on the 5-tuple $t$ that is equal to $c$ at coordinate $i$, and equal to $b$ elsewhere. We have that $p'(t)$ is equal to $\mathsf{mid}(\{1, 2, 3, 4, 5\}) = 3$ or $\mathsf{mid}(\{1, 2, 3, 4, 5\} \setminus \{i\})$ depending on whether or not $s(b) = 0$; in either situation, $p'(t)$ is not equal to $i$, and the claim follows.

We now consider the case $i = 3$. Let $b \in B$ be an element; we want to show that $b$ is in the image of $p_3^c$. If $s(b) = 1$, then we show that $p_3^c(b, b, b, b) = b$. We have $p_3^c(b, b, b, b) = p(b, b, c, b, b)$. Observe that the value $p'(b, b, c, b, b)$ is equal to $\mathsf{mid}(\{1, 2, 4, 5\}) = 2$, so $p(b, b, c, b, b) = b$. If $s(b) = 0$, we argue as follows. Fix $d$ to be an element of $B_1$. We show that $p_3^c(b, b, b, d) = b$. We have $p_3^c(b, b, b, d) = p(b, b, c, b, d)$. The value $p'(b, b, c, b, d)$ is equal to $\mathsf{mid}(\{1, 2, 3, 4\}) = 2$, so $p(b, b, c, b, d) = b$. □

*Example 3.* (Disconnected structures) Let us say that a structure $\mathbf{B}$ on signature $\sigma$ is *disconnected* if there exists a partition $B_0 \cup B_1$ of $B$ composed of two non-empty sets such that for each symbol $R \in \sigma$ and for each tuple $(b_1, \ldots, b_k) \in R^{\mathbf{B}}$, it holds that either $\{b_1, \ldots, b_k\} \subseteq B_0$ or $\{b_1, \ldots, b_k\} \subseteq B_1$.

Let $\mathbf{B}$ be a disconnected structure. We make use of the operations defined in the previous example (Example 2), but assume now that they are with respect to a partition $B = B_0 \cup B_1$ that witnesses the disconnectivity of $\mathbf{B}$. We show that the operation $p : B^5 \to B$ is a polymorphism of $\mathbf{B}$. By the discussion on collapsibility in the previous example, this yields that $\mathsf{QCSP}(\mathbf{B})$ reduces to $\mathsf{CSP}(\mathbf{B}^*)$. Let $(b_1^1, \ldots, b_k^1), \ldots (b_1^5, \ldots, b_k^5)$ be tuples in a relation $R^{\mathbf{B}}$. We want to show that $(p(b_1^1, \ldots, b_1^5), \ldots, p(b_k^1, \ldots, b_k^5)) \in R^{\mathbf{B}}$. We claim that $p'(b_1^1, \ldots, b_1^5) = \cdots = p'(b_k^1, \ldots, b_k^5)$, which suffices. For each $i$ (from 1 to 5), we have $s(b_1^i) = \cdots = s(b_k^i)$. For values of $j$ ranging from 1 to $k$, the values $w(s(b_j^1), \ldots, s(b_j^5))$ are all equal, and the sets $\{i \mid s(b_j^i) = w(s(b_j^1), \ldots, s(b_j^5))\}$ are all equal; the claim follows. □

*Remark 7.* Completing the complexity classification of the QCSP on undirected graphs, which was initiated in [28], is an open issue. A complementary objective that we believe could be of interest is to understand which algebras coming from graphs have the PGP.

As an analog of Conjecture 3 in the unbounded alternation setting, we conjecture that here, lack of the PGP property implies PSPACE-completeness of the QCSP. Again, we form a conjecture only on structures of the form $\mathbf{B}^*$.

*Conjecture 6.* Let $\mathbf{B}$ be a finite relational structure on a finite signature. If the algebra $\mathbb{A}_{\mathbf{B}}$ does not have the PGP property, then the problem $\mathsf{QCSP}(\mathbf{B}^*)$ is PSPACE-complete.

*Remark 8.* A non-trivial PSPACE-completeness result for a problem $\mathsf{QCSP}(\mathbf{B})$ can be found in [6, Section 6.2]. In particular, such a complexity result is given for certain structures preserved by semilattices without an identity element (recall Example 1). Identifying general sufficient conditions for the PSPACE-hardness of $\mathsf{QCSP}(\mathbf{B})$ is an issue for future research.

Our conjectures focus on structures having constants, that is, structures having the form $\mathbf{B}^*$. We do not yet dare form conjectures on general structures! On a structure $\mathbf{B}^*$, when one has a reduction from $\mathsf{QCSP}(\mathbf{B}^*)$ to $\mathsf{CSP}(\mathbf{B}^*)$, these two problems are interreducible and have the same complexity (as $\mathsf{CSP}(\mathbf{B}^*)$ is a special case of and reduces to $\mathsf{QCSP}(\mathbf{B}^*)$). Note that the conjectures in this section, along with the conjecture that the CSP admits a dichotomy, predict a P/NP-complete/PSPACE-complete trichotomy in the complexity of the problems $\mathsf{QCSP}(\mathbf{B}^*)$: in the presence of the PGP property, the problem $\mathsf{QCSP}(\mathbf{B}^*)$ is predicted to be interreducible with $\mathsf{CSP}(\mathbf{B}^*)$ (Conjecture 5); otherwise, the problem $\mathsf{QCSP}(\mathbf{B}^*)$ is predicted to be PSPACE-complete (Conjecture 6).

In Examples 2 and 3, we saw structures for which collapsibility can be used to give a reduction from $\mathsf{QCSP}(\mathbf{B})$ to $\mathsf{CSP}(\mathbf{B}^*)$. We now show, via a concrete example, that when $\mathbf{B}$ does not have constants, it is possible that there is such a reduction but that $\mathsf{CSP}(\mathbf{B}^*)$ does *not* characterize the complexity of $\mathsf{QCSP}(\mathbf{B})$: in passing from the problem $\mathsf{QCSP}(\mathbf{B})$ to $\mathsf{CSP}(\mathbf{B}^*)$, one can see a jump in complexity.

*Example 4.* (Jump from $\mathsf{QCSP}(\mathbf{B})$ to $\mathsf{CSP}(\mathbf{B}^*)$) Let $\sigma$ be the signature $\{S\}$ where $S$ is of arity 3. Let $\mathbf{A}$ be the structure on $\sigma$ with universe $\{0,1\}$ and where $S^\mathbf{A} = \{(a,b,c) \in \{0,1\}^3 \mid a = b \text{ or } b = c\}$. Let $\mathbf{B}$ be the structure on $\sigma$ with universe $\{0,1,2\}$ and where $S^\mathbf{B} = S^\mathbf{A} \cup \{(2,2,2)\}$. The structure $\mathbf{B}$ is clearly disconnected via the partition $B = \{0,1\} \cup \{2\}$, and hence the discussion in Example 3 implies that there is a reduction from $\mathsf{QCSP}(\mathbf{B})$ to $\mathsf{CSP}(\mathbf{B}^*)$. We show that the problem $\mathsf{QCSP}(\mathbf{B})$ is polynomial-time decidable, but the problem $\mathsf{CSP}(\mathbf{B}^*)$ is NP-complete.

We first show that the problem $\mathsf{CSP}(\mathbf{B}^*)$ is NP-complete. We reduce from the problem $\mathsf{CSP}(\mathbf{A}^*)$, which is NP-complete by Schaefer's theorem. Given an instance $\Phi = \exists v_1 \ldots \exists v_n \phi$ of $\mathsf{CSP}(\mathbf{A}^*)$, where $\phi$ is quantifier-free, the reduction creates the instance $\Phi'$ defined as

$$\exists v_1 \ldots \exists v_n \exists c_0 \exists c_1 (\phi \wedge R_0(c_0) \wedge R_1(c_1) \wedge \bigwedge_{i=1}^{n} S(c_0, v_i, c_1)).$$

In this latter instance, the variables $c_0$ and $c_1$ are assumed to be fresh variables that do not occur among $v_1, \ldots, v_n$; via the formulas $R_0(c_0)$ and $R_1(c_1)$, they are forced to the values 0 and 1. The formulas $S(c_0, v_i, c_1)$ force each variable $v_i$ to take on the value 0 or 1. (Note that this forcing can also be done with just one of the constants $c_0, c_1$; for instance, the formula $S(c_0, c_0, v_i)$ also forces the variable $v_i$ to take on the value 0 or 1.) With these facts in mind, it is readily verified that $\Phi$ is true on $\mathbf{A}^*$ if and only if $\Phi'$ is true on $\mathbf{B}^*$.

We now show that $\mathsf{QCSP}(\mathbf{B})$ is polynomial-time decidable. Given an instance $\Phi$ of $\mathsf{QCSP}(\mathbf{B})$, let $G_\Phi$ be the undirected graph whose vertices are the variables of $\Phi$ and where an edge $\{u,v\}$ is present if and only if $u$ and $v$ occur together in a predicate application. For each connected component $C$ of $G_\Phi$, we define $C'$ to be the set obtained by removing the variable that is quantified first in $C$. We claim that $\Phi$ is true if and only if the following condition holds: for each connected component $C$ of $G_\Phi$, the set $C'$ contains only existentially quantified variables. This suffices, since the condition is readily checkable in polynomial time.

We verify the claim as follows. We view $\Phi$ as a game between two players, universal and existential, that set the respectively quantified variables in the order given by the quantifier prefix; the universal player tries to falsify the quantifier-free part $\phi$ of $\Phi$, whereas the existential player tries to satisfy $\phi$. If the condition holds, then for each connected component $C$, by setting all of the variables in $C'$ appropriately, the existential player can guarantee that all variables in $C$ are set to the same value. This suffices to satisfy $\phi$, since all three constant tuples $(0,0,0), (1,1,1), (2,2,2)$ are contained in $S^{\mathbf{B}}$. If the condition does not hold, then there exists a connected component $C$ such that there is a universally quantified variable $y$ in $C'$. Let $v$ be the variable in $C$ but not in $C'$, and let us call $\{0,1\}$ and $\{2\}$ the *blocks* of $B$. The universal player can then guarantee that $y$ is set to a value that is in a different block from the value given to $v$. This suffices to spoil $\phi$, since to satisfy $\phi$ it must be that all variables in $C$ are set to values in the same block. □

From the perspective of this example, the approach of giving a reduction from $\mathsf{QCSP}(\mathbf{B})$ to $\mathsf{CSP}(\mathbf{B}^*)$ is not as fine a tool as one would like for understanding the complexity of $\mathsf{QCSP}(\mathbf{B})$: while this approach gives an NP upper bound on $\mathsf{QCSP}(\mathbf{B})$, the problem $\mathsf{CSP}(\mathbf{B}^*)$ does not always yield the more detailed information of whether or not $\mathsf{QCSP}(\mathbf{B})$ is in P. Broadly speaking, the development of tools for understanding the QCSP on general structures (by which is meant structures not necessarily having constants) would be very welcome.

**Acknowledgements.** Simone Bova and Moritz Müller provided numerous helpful suggestions. Johan Thapper supplied pointed feedback and many corrections that induced both revision of this article as well as further meditation. I also thank Andrei Bulatov, Florent Madelaine, Barny Martin, and Peter Mayr for useful discussions, and Matt Valeriote for his support. Finally, I express gratitude to Dexter Kozen, who supervised my doctoral thesis on quantified constraint satisfaction, which thesis is the ultimate origin of these meditations. The author is supported by the Spanish program "Ramon y Cajal" and MICINN grant TIN2010-20967-C04-02.